# Confinement-Controlled Morphology and Stability of One-Dimensional CrI$_3$ Nanotubes

*Ihsan Çaha\*, Aqrab ul Ahmad and Francis Leonard Deepak\**

INL - International Iberian Nanotechnology Laboratory,

Avenida Mestre José Veiga s/n,

Braga 4715-330,

Portugal

**Corresponding Author**

E-mail: ihsan.caha@inl.int, leonard.francis@inl.int

ABSTRACT

Integrating monolayers derived from 2D van der Waals (vdW) magnetic materials into next-generation technological applications remains a significant challenge due to their structural and magnetic instability issues. Template-assisted encapsulation is a potential route for the growth of stable 2D monolayers aimed at designing novel 1D heterostructures, opening new avenues for studying low-dimensional quantum effects and spin-related phenomena. In this study, we explored the diameter-dependent encapsulation of 2D $CrI_3$ crystals using multi-walled carbon nanotubes (MWCNTs) as nanoscale host templates. Advanced microscopic analysis revealed distinct structural transitions, ranging from internal nanorod encapsulation to external shell formation, directly influenced by the host nanotube diameter (2–20 nm). Furthermore, statistical analysis of structural morphologies indicates that $CrI_3$ nanorods preferentially form within MWCNTs with inner diameters up to ~5 nm, while single-walled $CrI_3$ nanotubes are stabilized in CNTs with diameters up to ~8 nm. For host CNTs exceeding ~10 nm in diameter, $CrI_3$ predominantly forms surface coatings rather than confined one-dimensional structures. In situ electron beam irradiation demonstrates the superior structural stability of single-walled $CrI_3$ confined within MWCNTs, while externally coated $CrI_3$ undergoes decomposition into metallic Cr clusters. Prolonged irradiation induces a morphological transformation of $CrI_3$ nanotubes into nanorods. These insights lay the groundwork for engineering robust, tunable 1D magnetic heterostructures of $CrI_3$ for spintronic and data storage applications.

1. INTRODUCTION

The environmental instability of emerging two-dimensional (2D) magnetic materials such as $CrI_3$, $CrCl_3$, $CrBr_3$, and $VI_3$ remains a major hurdle for their integration into next-generation electronic and quantum systems.[1–4] Encapsulation within inert organic and inorganic van der Waals materials like hexagonal boron nitride (h-BN) has proven effective in mitigating ambient degradation.[5] However, this strategy imposes practical limitations: encapsulation typically follows device fabrication, which constrains design flexibility and hinders large-scale integration.[6,7] An alternative and promising approach involves the encapsulation of these air-sensitive 2D materials within one-dimensional (1D) hollow-core nanotubes, such as carbon nanotubes (CNTs) or boron nitride nanotubes (BNNTs).[8] These nanoconfined environments offer robust environmental shielding and promote direct dimensional reduction, enabling the construction of novel 1D magnetic heterostructures.[9] The resulting dimensional confinement enhances spin fluctuations, disrupts conventional magnetic ordering, and gives rise to emergent phenomena such as increased magnetic anisotropy.[10] These effects are critical for advancing spintronic devices, high-density magnetic storage, and nanoscale sensing technologies. Therefore, despite growing interest in low-dimensional magnetic systems, exploring true one-dimensional (1D) polymorphs, particularly single-walled magnetic nanotubes, remains scarce. While some quasi-1D magnetic materials have been modestly investigated, the controlled synthesis of fully isolated, truly 1D single-walled magnetic phases within confined templates remains an unresolved challenge.[9,11]

In recent years, considerable attention has been directed toward utilizing carbon nanotubes particularly single-walled carbon nanotubes (SWCNTs) as one-dimensional (1D) confinement platforms for tailoring the structure and properties of two-dimensional (2D) materials at the nanoscale. A wide range of encapsulation studies has been carried out with materials such as $PbI_2$, $BiI_3$, $GdI_3$, and $MoS_2$, which are known for their intrinsic 2D layered structures.[12–16] However, the potential application of multi-walled carbon nanotubes (MWCNTs) for this type of filling or coating has not been thoroughly investigated. Gaining a deep understanding of how MWCNTs can serve as nanoscale templates for the self-assembly and confinement of 2D magnets is essential for realizing their full capabilities in future applications, especially in the areas of spintronics, magnetic nanodevices, and quantum technologies.[17] $CrI_3$, a prototypical 2D ferromagnetic insulator, is particularly vulnerable to ambient degradation via hydrolysis and oxidation, producing by-products such as $Cr(OH)_3$, $I_2$, $CrO_3$, and $HI$.[18] This reactivity severely limits its functional deployment in devices and underscores the need for robust encapsulation strategies.[18] Despite its importance, the encapsulation of $CrI_3$ in CNTs remains insufficiently explored, and a comprehensive understanding of its structural evolution and environmental stability under 1D confinement is still lacking.

In this study, we investigate the encapsulation of $CrI_3$ within MWCNTs of varying diameters to explore the structural confinement limits of $CrI_3$ and the influence of dimensional reduction on its morphology and crystallinity. We demonstrate the formation of $CrI_3$ nanostructures ranging from single-walled nanotubes to nanorods with high filling efficiency. Structural characterization using high-resolution transmission electron microscopy (HRTEM) and high angle annular dark-field scanning transmission electron microscopy (HAADF-STEM) confirms the integrity and uniformity of the encapsulated $CrI_3$ within the nanotubes. Moreover, our findings indicate a strong diameter dependence in forming single-walled $CrI_3$ nanotubes, with optimal confinement required for their structural realization. In situ TEM analysis further reveals that these encapsulated $CrI_3$ nanotubes remain stable under electron beam exposure, in contrast to $CrI_3$ layers deposited outside of host MWCNTs, which undergo beam-induced decomposition into metallic chromium (Cr). These observations highlight the critical role of spatial confinement in preserving the structural phase and preventing reduction in Cr-halide-based 1D magnetic heterostructures. It is anticipated that establishing a comprehensive framework for understanding template-directed growth, diameter-dependent phase stability, and structural transformations of $CrI_3$ within carbon nanotubes will lay the groundwork for the development of $CrI_3$-based next-generation one-dimensional quantum and spintronic devices.

2. RESULTS AND DISCUSSION

**Figure 1a** illustrates the synthesis and formation mechanism of one-dimensional (1D) $CrI_3$ nanostructures within MWCNTs via capillary filling. First, $CrI_3$ and MWCNTs are mixed and sealed in a quartz ampoule under vacuum. This ampoule is then positioned in a two-zone horizontal furnace for further heat treatment. Upon heating above the melting point of $CrI_3$ (~650 °C), the compound transitions to a molten state and infiltrates the cavities of the MWCNTs via capillary action. This infiltration is stabilized by strong van der Waals interactions between the $CrI_3$ layers and the graphitic walls of the MWCNTs, promoting confinement within or along the nanotube surface which acts as a template. Similar molten-phase infiltration processes have been successfully employed for layered materials such as $PbI_2$[12,19–21], $GdI_3$[14] and $BiI_3$[15], leading to well-ordered confined structures. This synthesis involves a cyclic thermal treatment (refer to Figure S1), alternating between high (650 °C) and intermediate (550 °C) temperature stages to optimize infiltration dynamics and promote crystallinity. After approximately 36 hours, this process ensures uniform filling and surface coating of $CrI_3$ within or along the MWCNTs. The final morphology and structural characteristics are strongly influenced by the interaction between the molten $CrI_3$ crystals wetting properties, the diameter of the MWCNTs, and the thermodynamic stability of the confined phase.[10]

**Figure 1b** provides direct evidence of CrI$_3$ encapsulation within MWCNTs. Based on prior studies of confined halide growth, such as the stepwise transformations described by Botos *et al.*,[22] the formation of the single-walled CrI$_3$ nanotube is likely to proceed through a sequential process. In this mechanism, molten CrI$_3$ initially infiltrates the CNT cavity and nucleates at discrete sites, forming short crystalline domains or intermediate fragments. These fragments subsequently undergo lateral growth and alignment, gradually connecting and reorganizing into a continuous single-walled tubular structure.[23] The CNT confinement not only restricts the growth dimensionally but also promotes this coalescence pathway by enforcing a common cylindrical geometry.[24] Such a confined, multistep formation route is characteristic of metal halides synthesized within nanotubes and highlights the nanoreactor role of MWCNTs in directing both structure and assembly of the encapsulated CrI$_3$ phase.

**Figure 1c** shows the formation of a CrI$_3$ single-walled nanotube coating on the external surface of a MWCNT. This coating morphology typically emerges when the CNT diameter is sufficiently large to reduce capillary pressure, hindering encapsulation and instead favoring external nucleation. The coating process resembles the conformal growth observed in transition metal dichalcogenide-based coaxial nanotubes, where an outer wall forms via van der Waals epitaxy along the surface of a core template, as demonstrated by Yomogida *et al.*[24] In our case, molten CrI$_3$ wets the CNT exterior and crystallizes as a single-walled tube that conforms to the host curvature. Furthermore, recent theoretical predictions on topological heteronanotubes suggest that such 1D coaxial systems can exhibit stacking-dependent symmetry breaking and interwall electronic coupling.[25] While our system differs in material composition, the underlying principle that the outer tube's structure and chirality can be guided or influenced by the template remains valid. These findings reinforce the notion that CNTs act as active growth scaffolds, not only dictating the geometry but potentially influencing the crystallographic orientation and electronic characteristics of CrI$_3$ coatings.

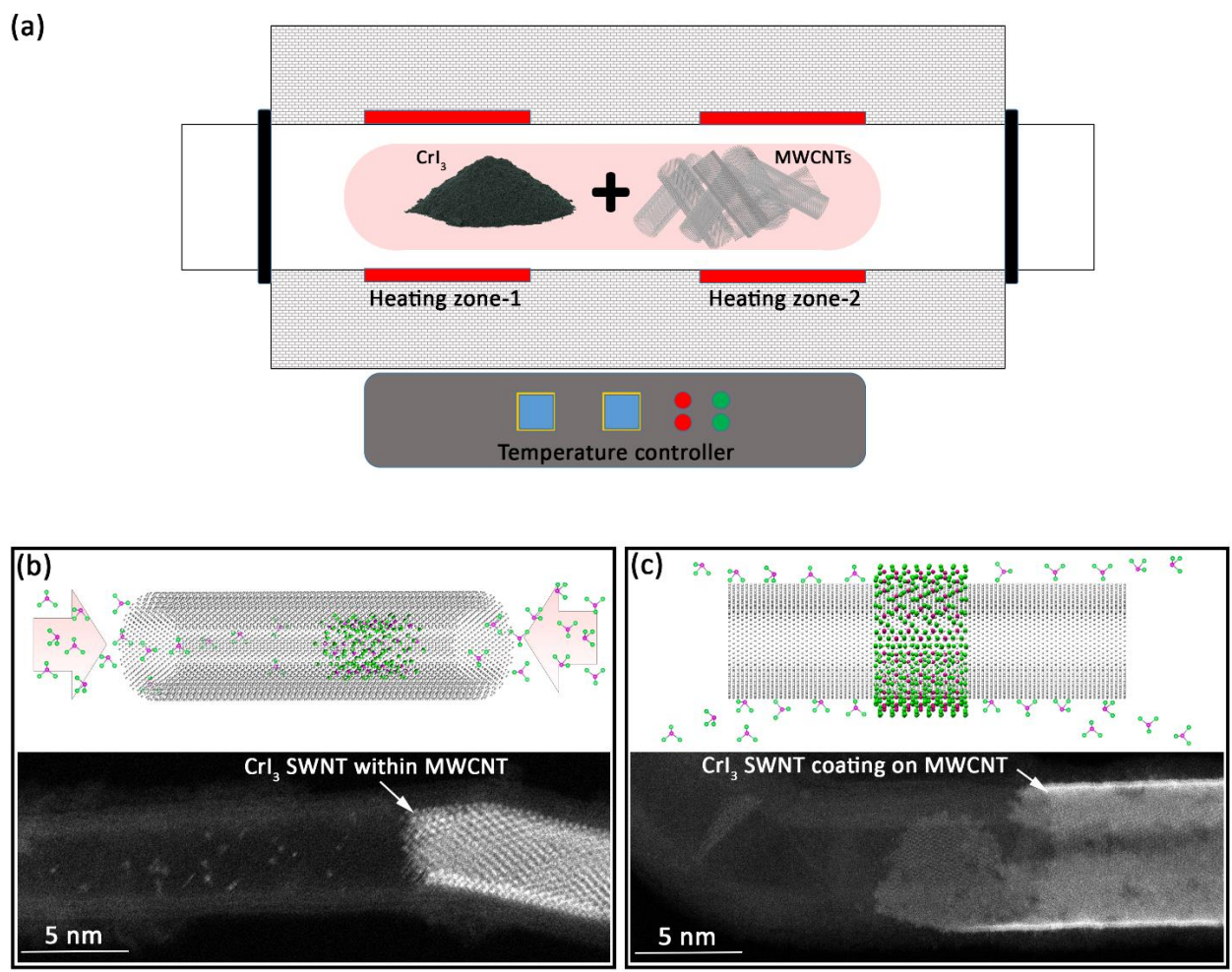

**Figure 1.** Synthesis and growth of 1D CrI$_3$ nanotubes in MWCNTs. (a) Schematic of the capillary filling process using a two-zone furnace setup. (b) Encapsulation process showing a HAADF-STEM image of CrI$_3$ forming a SWNT within a MWCNT. (c) Coating process illustrated by a HAADF-STEM image showing CrI$_3$ forming a SWNT coating on the external surface of a MWCNT.

**Figure 2** presents a detailed statistical analysis of CrI$_3$ structural formation within CNTs of varying diameters, highlighting how spatial confinement influences morphology. Panels (a–c) display histograms corresponding to encapsulated CrI$_3$ nanorods, single-walled CrI$_3$ nanotubes, and CrI$_3$ nanotube coatings, respectively. The measured distributions reveal a diameter-dependent transition: CrI$_3$ nanorods predominantly form in CNTs with inner diameters of 4.0 ± 1.0 nm, while single-walled CrI$_3$ nanotubes appear within slightly larger CNTs, averaging 5.6 ± 1.5 nm. Beyond ~10 nm, CrI$_3$ preferentially crystallizes as single-walled coatings on the external surfaces of MWCNTs with an average external tube diameter of 13.1 ± 2.2 nm rather than forming encapsulated nanotubes.

The diameter-dependent morphology strongly suggests that capillary-driven infiltration alone is insufficient to dictate $CrI_3$ structure, confinement effects play a dominant role. This trend aligns with previous studies on $PbI_2$ nanotube formation, where a critical diameter threshold was identified for nanotube versus nanorod formation.[12] Similar behavior has also been observed for other van der Waals materials confined in nanotubes, such as $MoS_2$[16,24] and $BiI_3$[15], where the interplay between template diameter, interfacial van der Waals interactions, and nucleation kinetics governed structure formation.

Theoretical calculations from our previous work[17] have further confirmed that metal halide nanotube formation is energetically favored within a specific diameter range, where wrapping energy remains moderate while encapsulation energy remains sufficiently high to stabilize the tubular phase. These results suggest that $CrI_3$ nanotubes form within an optimal CNT diameter range, where confinement provides sufficient structural stability for a tubular phase. When the available space is too small, $CrI_3$ instead crystallizes as a bulk-like nanorod, whereas larger diameters lead to external coating rather than encapsulated nanotube formation. This behavior mirrors previous observations in $PbI_2$, where nanotube formation was only achieved within a specific host diameter window.[12,19]

Panel (**d**) summarizes these findings in a morphological phase diagram, mapping the formation regions of each $CrI_3$ structure. Encapsulated nanorods dominate in the smallest-diameter CNTs, single-walled nanotubes form in an intermediate range, and coatings are favored in larger-diameter CNTs, reinforcing the critical influence of spatial confinement on $CrI_3$ crystallization pathways. These results highlight the potential of CNT diameter selection as a precise tool for engineering $CrI_3$-based 1D heterostructures with tailored structural and electronic properties.

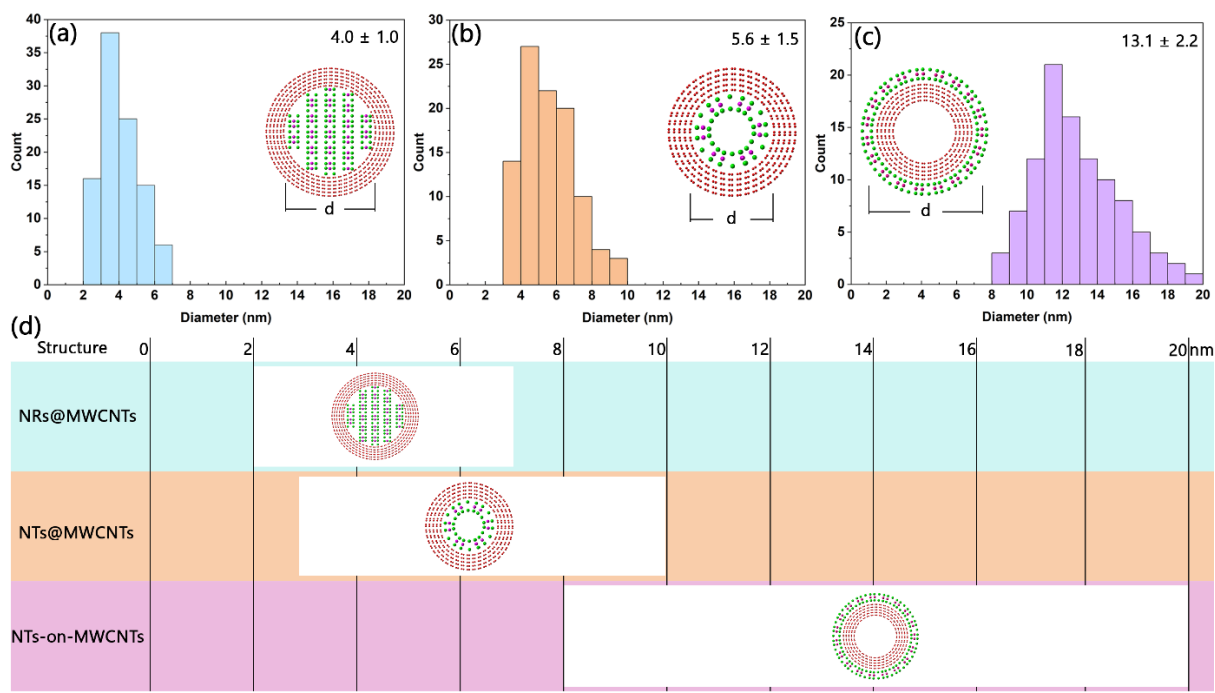

Figure 2. Statistical analysis of $CrI_3$ structural formation within CNTs of varying diameters. (a–c) Histograms showing the measured diameter distributions of $CrI_3$ nanorods encapsulated within MWCNTs (NRs@MWCNTs), single-walled $CrI_3$ nanotubes encapsulated within MWCNTs (NTs@MWCNTs), and $CrI_3$ single-walled nanotube coatings on the external surface of MWCNTs (NTs-on-MWCNTs), respectively. Insets illustrate the corresponding structural models. (d) A morphological phase diagram mapping the preferred formation regions of each $CrI_3$ structure as a function of CNT diameter. The transition from nanorods to single-walled nanotubes occurs within 4–6 nm, while $CrI_3$ coatings become dominant beyond 10 nm, highlighting the strong influence of confinement on $CrI_3$ crystallization pathways.

The influence of CNT confinement on $CrI_3$ morphology is explored in greater detail in **Figure 3,** which presents the different structural morphologies of $CrI_3$ that form depending on the diameter of the CNT host. The observed structures include encapsulated nanorods ($CrI_3$ NR@MWCNT), encapsulated single-walled $CrI_3$ nanotubes ($CrI_3$ SWNT@MWCNT), single-walled $CrI_3$ coatings on MWCNT surfaces ($CrI_3$ SWNT coating on MWCNT), and double-walled $CrI_3$ coatings ($CrI_3$ DWNT coating on MWCNT). The schematic cross-section representations in **Figure 3a** illustrate these distinct morphologies, while the corresponding top-view structural models in **Figure 3b** provide a visualization of their atomic arrangements. The HAADF-STEM images in **Figure 3c** further confirm the presence of these structures, with inset cross-sectional intensity profiles illustrating the material contrast and radial distribution of CrI$_3$ within or on the CNT walls.

The distribution of these structures correlates strongly with CNT diameter, indicating that while capillary forces contribute to CrI$_3$ infiltration, confinement effects play a decisive role in dictating final morphology. For narrow-diameter MWCNTs (2 to 10 nm), CrI$_3$ is predominantly encapsulated, forming either nanorods or single-walled nanotubes inside the CNTs. A nanorod refers to a bulk-like CrI$_3$ structure (a stack of 2D CrI$_3$ layers) confined within the CNT cavity, whereas a single-walled CrI$_3$ nanotube consists of a well-ordered, hollow tubular arrangement of CrI$_3$ atoms that conforms to the cylindrical geometry of the CNT. The presence of CrI$_3$ nanorods suggests that, in some cases, CrI$_3$ retains a more compact structure rather than forming a nanotube. This behavior may arise due to factors such as crystallization kinetics, suboptimal wetting interactions with CNT interiors, or structural imperfections in the CNTs. Blockages or irregularities within the CNT may locally restrict CrI$_3$ infiltration, leading to the accumulation of bulk-like material instead of a uniform nanotube. Similar deviations have been observed in other encapsulated van der Waals materials, such as CeI$_3$, ZnI$_2$, GdI$_3$, PbI$_2$, and BiI$_3$, where CNT structure plays a crucial role in determining final morphology.[12,14,15,23]

To further confirm the successful encapsulation of CrI$_3$ within CNTs, electron energy-loss spectroscopy (EELS) and energy-dispersive X-ray spectroscopy (EDS) mapping was performed. The EELS spectroscopy and elemental distribution across an encapsulated CrI$_3$ nanotube within CNTs is presented in **Figure 2d**, where the measured C $K$, Cr $L_{2,3}$, and I $M_{4,5}$ edges with their EELS maps confirm, the successful encapsulation of a single-walled nanotube. Furthermore, EDS elemental mapping of an encapsulated nanotube and nanorod are given in **Figure S2**, where the encapsulated CrI$_3$ nanorods and single-walled CrI$_3$ nanotubes confirm the presence of Cr and I elements along the CNT axis, verifying that the observed structures correspond to CrI$_3$ rather than contamination or amorphous deposits.

In contrast, larger-diameter CNTs (10 to 20 nm) show a significant shift in CrI$_3$ morphology, favoring surface coatings rather than full encapsulation. In these cases, the capillary pressure is lower, reducing the driving force for CrI$_3$ infiltration. As a result, CrI$_3$ tends to nucleate along the CNT exterior, forming either a uniform single-walled layer (CrI$_3$ SWNT coating on MWCNT) or, in rare instances, an additional secondary layer (CrI$_3$ double-walled nanotube (DWNT) coating on MWCNT). A single-walled CrI$_3$ coating refers to a monolayer of CrI$_3$ atoms covering the CNT surface, while a double-walled $CrI_3$ coating consists of two distinct layers of $CrI_3$ stacking on the CNT exterior, suggesting a possible layer-by-layer growth mechanism. The formation of such coatings is consistent with previous observations of metal halide deposition on CNTs, where weak interlayer interactions allow for multilayer stacking.[24,26] The fact that encapsulated $CrI_3$ nanotubes are rarely observed in larger-diameter CNTs further supports the idea that strong confinement is required to induce the nanotube morphology.

These results indicate that $CrI_3$ morphology is not only determined by the capillary wetting process but also by the degree of confinement imposed by the CNT template. Smaller CNT diameters promote full encapsulation, leading to stable nanotubes, while larger CNT diameters favor the nucleation of external coatings rather than tubular encapsulation, underscoring the importance of spatial confinement in guiding $CrI_3$ self-assembly.

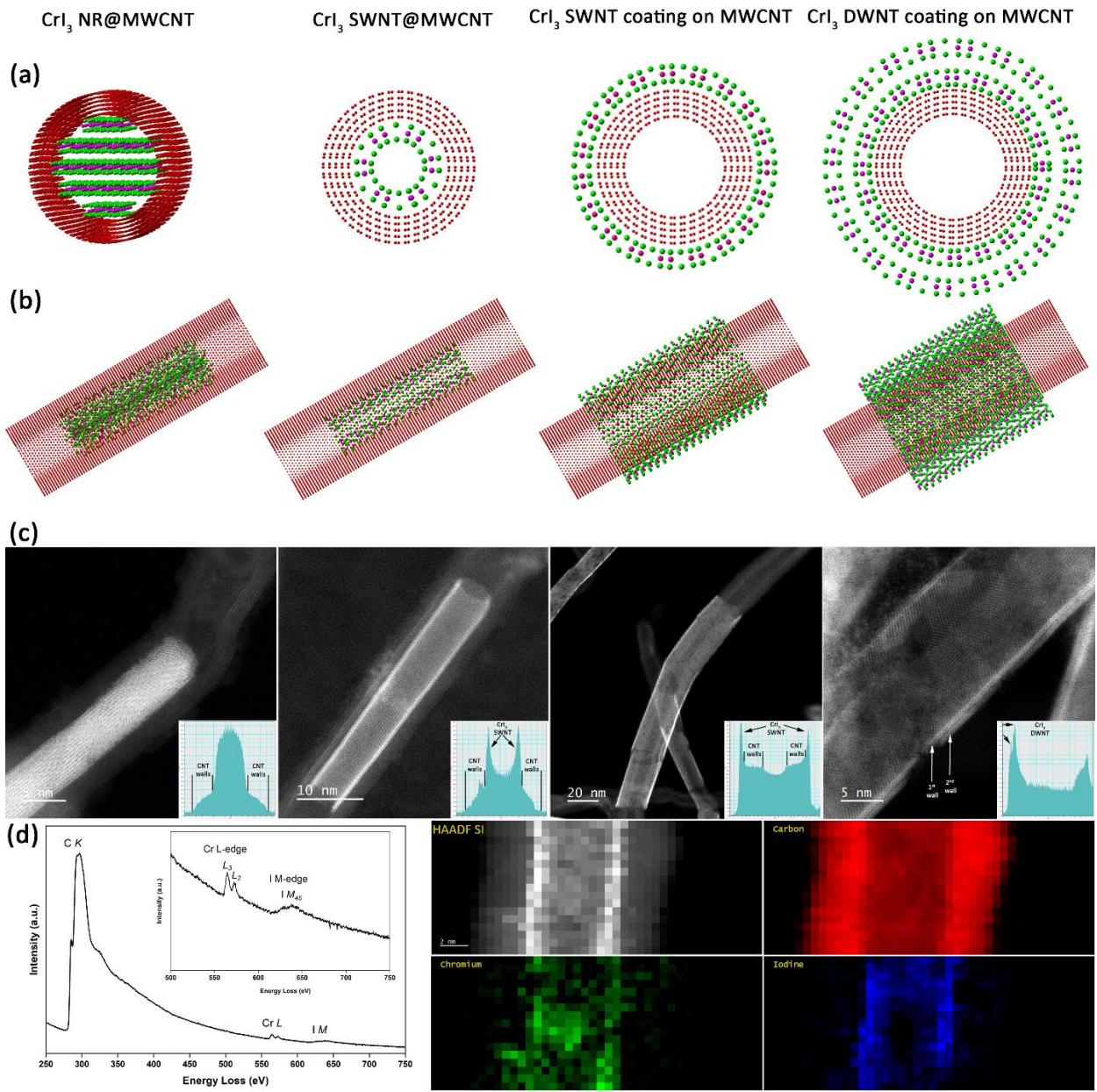

**Figure 3.** Structural morphologies of CrI$_3$ formed within and around MWCNTs. (a) Schematic cross-sectional illustrations of the four observed configurations: CrI$_3$ nanorods encapsulated in MWCNTs (CrI$_3$ NR@MWCNT), single-walled CrI$_3$ nanotubes encapsulated in MWCNTs (CrI$_3$ SWNT@MWCNT), single-walled CrI$_3$ coatings on MWCNT surfaces (CrI$_3$ SWNT coating on MWCNT), and double-walled CrI$_3$ coatings (CrI$_3$ DWNT coating on MWCNT). (b) 3D top-view structural models showing atomic arrangements for each case. (c) Corresponding HAADF-STEM images, with inset line profiles demonstrating intensity variations across the CNT diameter,

confirming the morphology and structural distinctions. (d) EELS spectrum, HAADF-STEM SI together with EELS mapping for C $K$, Cr $L_{2,3}$, and I $M_{4,5}$ edges.

Low-magnification HAADF-STEM images in **Figure S3** provide an overview of the large-area distribution and uniformity of CrI$_3$ nanostructures within and on MWCNTs. Panel (a) highlights a dense network of MWCNTs containing encapsulated CrI$_3$, while panel (b) reveals CrI$_3$-coated MWCNTs with similarly high structural uniformity. In both cases, the CrI$_3$ morphologies appear well-aligned along the CNT axes, suggesting that the CNT template effectively guides the formation of both encapsulated and coated configurations.

Building upon these observations, **Figure 4** presents high-resolution imaging and structural analysis of individual CrI$_3$ nanotubes. Panel (a) shows a representative HAADF-STEM image of a CrI$_3$-coated MWCNT, where the coating exhibits excellent crystallinity and continuity. The FFT inset confirms the presence of zigzag stacking in the CrI$_3$ layer, which may arise from strain relaxation mechanisms or thermodynamically favorable growth directions, as previously reported by Kuklin *et al.*[27] However, CrI$_3$ coatings exhibit significant instability under electron beam exposure, as demonstrated in panel (b). With an electron dose of ~0.5 × 10$^4$ e$^-$ Å$^{-2}$, rapid degradation of the CrI$_3$ coatings was observed, leading to the formation of metallic chromium clusters, confirmed by HRTEM and FFT analysis in panel (c). The instability of CrI$_3$ under electron beam irradiation aligns with previous studies, which report that CrI$_3$ degrades much faster than other air-sensitive 2D materials, such as black phosphorus, NbSe$_2$, InSe, and CrGeTe$_3$.[18] This degradation is attributed to electron-induced knock-on damage and ionization effects, which result in the loss of iodine and the subsequent reduction of CrI$_3$ into metallic Cr. Similar degradation pathways have been observed in 2D CrI$_3$ flakes, where exposure to ambient conditions rapidly leads to hydrolysis and oxidation transformation of CrI$_3$ coatings under beam exposure suggesting that encapsulation strategies, such as hBN coating, could be necessary to enhance the long-term stability of CrI$_3$ nanotubes.[6]

Regarding encapsulated CrI$_3$ nanotubes, panels d-f provide insight into their structural and chemical response to electron beam exposure. Prior to irradiation (panel 4d), the encapsulated single-walled CrI$_3$ nanotube exhibits a well-defined hollow tubular morphology with high contrast and uniform thickness, indicating its structural integrity and crystalline order. After 365 seconds of continuous electron beam exposure at ~0.5 × 10$^4$ e$^-$ Å$^{-2}$ (panel 4e), the nanotube undergoes a distinct transformation into a denser, rod-like structure. This morphological evolution suggests that, although the encapsulated CrI$_3$ does not experience compositional degradation, it undergoes a beam-driven structural rearrangement. This is further supported by the STEM-EDS elemental map in panel 4f, which confirms the uniform retention of Cr and I, underscoring that the transformation is structural rather than chemical in nature, a key distinction from the degradation behavior observed in the coating morphology.

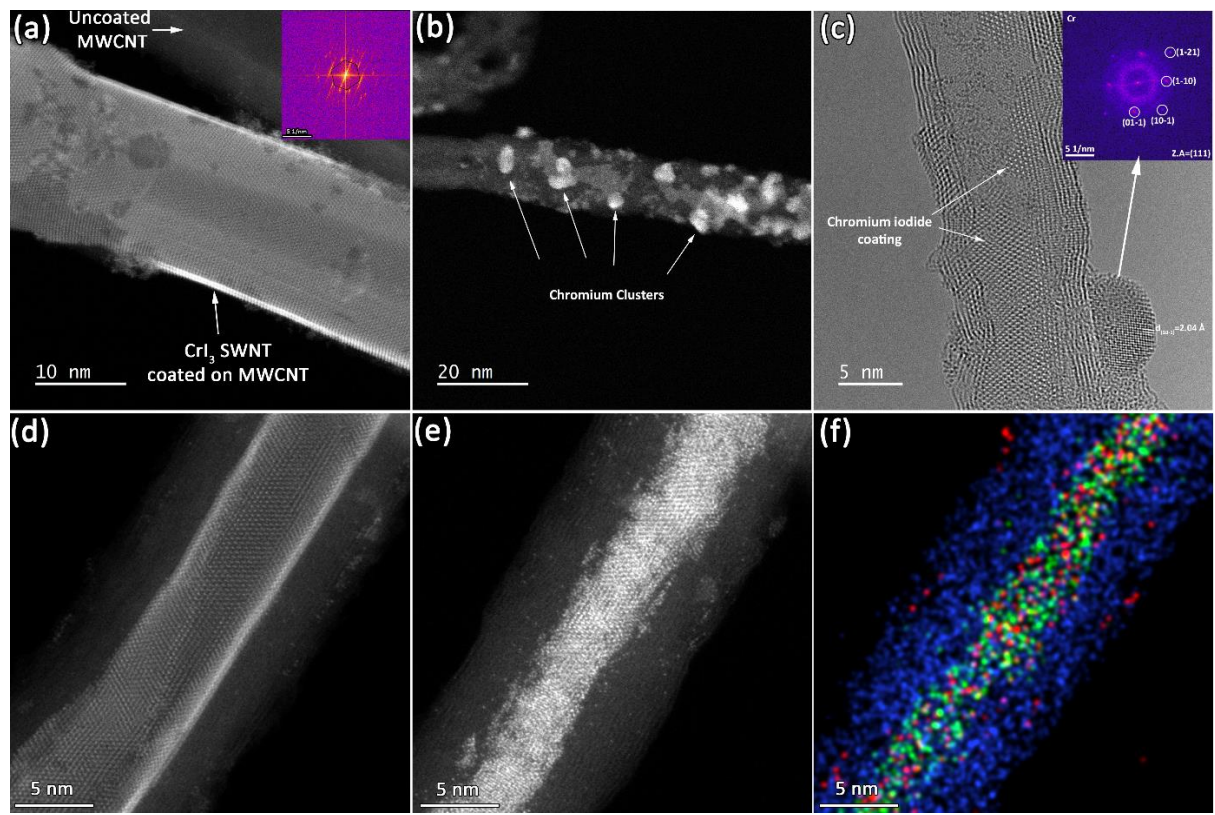

**Figure 4.** Structural, crystallographic, and stability characterization of CrI$_3$ coatings and encapsulated nanotubes within MWCNTs. (a) A HAADF-STEM image of CrI$_3$-coated MWCNTs, with FFT inset confirming zigzag stacking. (b) HAADF-STEM image showing chromium clustering under immediate beam irradiation. (c) HRTEM image of CrI$_3$ coating with FFT inset indicating metallic Cr formation. (d and e) HAADF-STEM images of encapsulated CrI$_3$ nanotube before and after 365 s of ~$0.5 \times 10^4$ e$^-$ Å$^{-2}$ beam exposure, respectively. (f) Corresponding STEM-EDS elemental map confirming uniform Cr and I distribution post-irradiation.

The structural evolution of single-walled CrI$_3$ nanotubes encapsulated within MWCNTs under electron beam irradiation is presented in **Figure 5**, where an in situ HRTEM study captures a dynamic transition from a CrI$_3$ nanotube to a nanorod-like morphology over approximately 90 seconds, at an electron dose of ~$0.5 \times 10^4$ e$^-$ Å$^{-2}$. This transformation suggests a restructuring of the CrI$_3$ lattice, potentially influenced by atomic migration, defect formation, or radiation-induced relaxation effects.[28] A similar transformation is observed in Supporting **Figure S5**, where an in situ study of a nanotube-nanorod junction under identical beam conditions demonstrates the same morphological transition, but at a faster rate, completed in ~70 seconds. This acceleration suggests that the presence of a pre-formed nanorod segment may act as a structural seed or strain focal point, facilitating beam-induced lattice reorganization.

Unlike structural collapse due to amorphization, the transition from nanotube to nanorod appears to be a well-defined reconfiguration of the CrI$_3$ lattice, likely governed by interlayer rearrangements and the intrinsic instability of the nanotube form under electron irradiation. The complete transformation sequence is provided in **Video S1**, where the nanotube progressively contracts and densifies into a nanorod, demonstrating a clear evolution in real-time under beam exposure.

In contrast to previous in situ studies on PbI$_2$ nanotubes encapsulated within CNTs, where beam irradiation facilitated the dissolution of a nanorod into a continuous nanotube, the transformation observed in CrI$_3$ follows the reverse trend, suggesting a fundamentally different response to electron beam exposure. In the case of PbI$_2$,[12] the migration of lead and iodine atoms led to the extension of the tubular structure, effectively filling an empty void within the nanotube, whereas in CrI$_3$, atomic displacement appears to favor the densification of the structure, causing the nanotube to evolve into a nanorod. This difference in behavior could be attributed to the stronger interlayer interactions within CrI$_3$ compared to PbI$_2$, as well as differences in defect mobility and iodine loss rates under electron irradiation. Given that the observed transition from nanotube to nanorod occurs within an encapsulated environment, the driving force behind this transformation is more likely associated with electron-beam-induced atomic rearrangements rather than chemical degradation.

Several mechanisms may contribute to this transformation, including interlayer collapse, beam-stimulated atomic migration, and structural relaxation effects.[29,30] The gradual conversion of the nanotube into a nanorod suggests that CrI$_3$ undergoes an interlayer reorganization, where the weakly bonded tubular walls contract and merge into a more compact rod-like structure, reducing the surface energy. Similar effects have been observed in other confined materials where external stimuli, such as electron irradiation or heating, drive structural rearrangements within CNTs.[12,30–32] Additionally, given the previously discussed instability of CrI$_3$ coatings under irradiation (**Figure 4**), knock-on effects and ionization-induced displacement may contribute to the progressive densification of the nanotube. The fact that the transformation is consistently observed in different nanotube segments further supports the idea that the nanotube configuration represents a metastable phase that readily collapses into a more thermodynamically stable nanorod structure when subjected to sufficient external energy. The full transformation process captured over an extended period is provided in **Video S2**, where the gradual progression of the structural rearrangement further validates the robustness of this phenomenon.

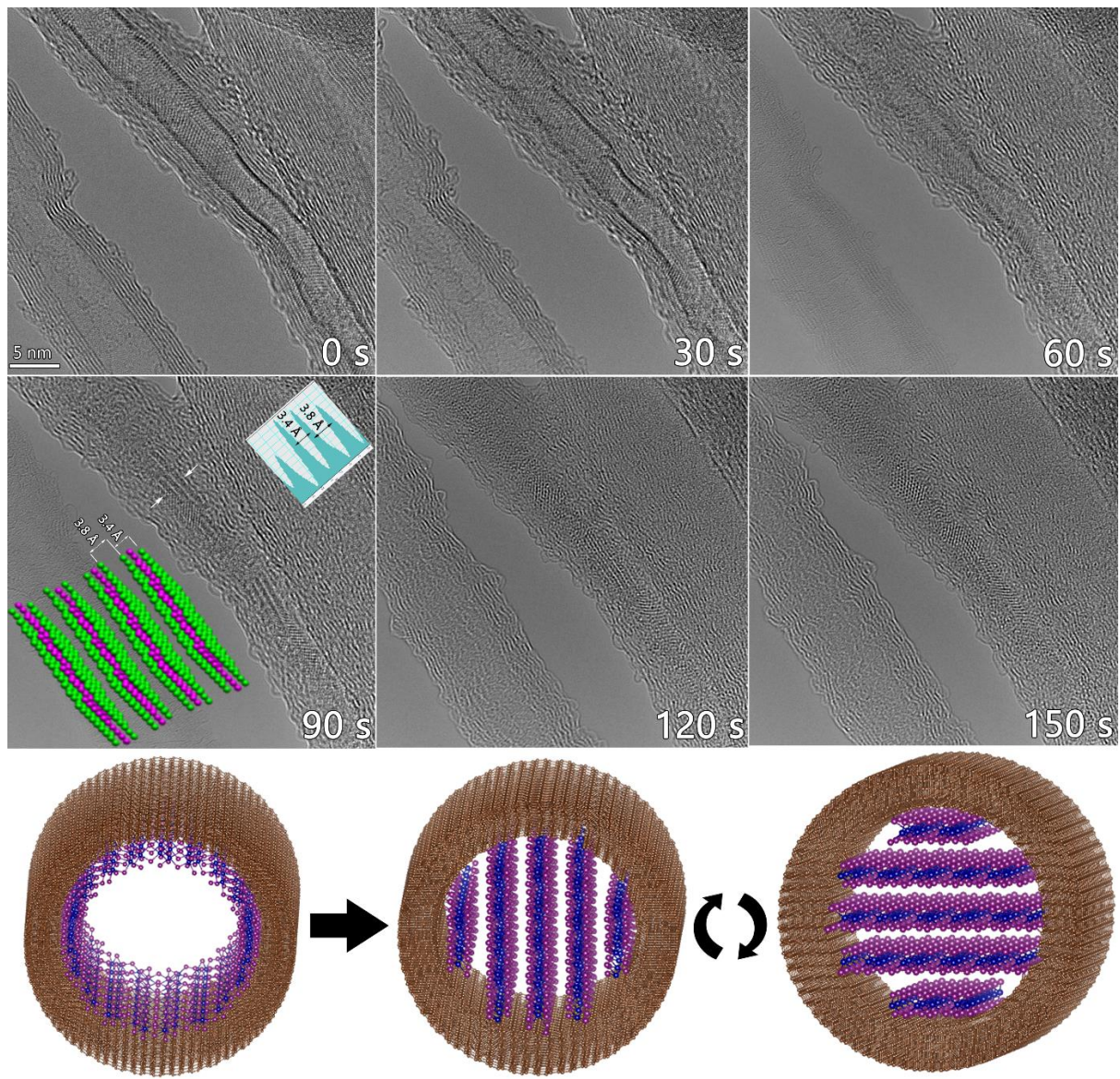

**Figure 5.** In situ TEM study of the electron beam-induced transformation of encapsulated CrI$_3$ nanotubes into nanorods. Sequential HRTEM images show the evolution of a single-walled CrI$_3$ nanotube confined within an MWCNT under an electron dose of ~$0.5 \times 10^4$ e$^-$ Å$^{-2}$, recorded over 150 seconds. The transition from a nanotube to a nanorod-like morphology occurs progressively, with significant structural rearrangement observed after 30 seconds. The bottom row presents a schematic representation of the transformation mechanism, illustrating the contraction and densification of the $CrI_3$ nanotube. The full in situ transformation process is provided in Video S1.

3. CONCLUSIONS

In this work, we establish that encapsulation within MWCNTs offers a powerful route to control the morphology, orientation, and crystallinity of $CrI_3$, enabling the formation of diverse one-dimensional van der Waals (1D vdW) heterostructures, including nanorods, surface coatings, and fully confined nanotubes. The observed morphological evolution is strongly governed by the diameter of the MWCNT hosts, revealing a critical confinement threshold necessary for stabilizing $CrI_3$ nanotubes. Furthermore, in situ TEM investigations uncover the dynamic and fragile nature of $CrI_3$, where electron beam exposure induces structural transitions from nanotubes to nanorods and drives the decomposition of coatings into metallic Cr clusters. These findings offer fundamental insights into the size and environment-dependent stability of vdW nanostructures and establish design principles for engineering robust 1D $CrI_3$ systems for next-generation quantum and spintronic technologies.

4. METHODS

*Synthesis of $CrI_3$ in Carbon Nanotubes*: Single-walled $CrI_3$ nanotubes were synthesized using a capillary infiltration method, employing multi-walled carbon nanotubes (MWCNTs) as templates with small-diameter distributions (2–10 nm) and large-diameter distributions (10–20 nm). Open-ended MWCNTs were used as received, while $CrI_3$, a hygroscopic and air-sensitive material with a melting point above 600°C, was handled under an argon atmosphere to prevent degradation. A homogeneous mixture of $CrI_3$ and MWCNTs was prepared in an argon-filled glovebox using an agate mortar and pestle and sealed in a quartz ampoule, which was evacuated to ~0.1 Pa to remove residual gases. The ampoule was then heated at 5°C/min to 650°C and maintained for 12 hours (Figure S1), allowing molten $CrI_3$ to infiltrate the CNTs via capillary forces. Subsequent controlled cooling to room temperature facilitated the stabilization of $CrI_3$ within the nanotube cavities and on their outer surfaces.

Upon completion, the ampoule was opened in an inert environment, and the $CrI_3$-filled and coated nanotubes were collected and stored under dry, oxygen-free conditions for further characterization.

*Structural and Morphological Characterization*: The morphology and structural characteristics of the $CrI_3$-filled and coated MWCNTs were analyzed using high-resolution transmission electron microscopy (HRTEM), and scanning transmission electron microscopy (STEM). HRTEM/STEM imaging was conducted on a FEI Titan Themis 60–300 kV microscope, equipped with both image and probe aberration correctors. For elemental composition and spatial distribution, EDS and EELS analyses were acquired with the same microscope (operated at 200 kV) equipped with a Gatan Enfinium dual-EELS spectrometer and a Super-X EDS detector. EELS spectrum acquisition was carried out by using 0.25 eV/channel dispersion and a 2.5 mm detector aperture. The full-width half maximum of the zero-loss peak was measured to be 0.4 eV without monochromation, which determines the energy resolution of the obtained spectra. EELS spectrum was captured by using a 55 pA electron beam, 1 s exposure time for 50 frames, which summed together, enables the reconstruction of a single spectrum. The background of the EELS spectra was subtracted by using the power-law method, and elemental mapping was employed by using Gatan's DigitalMicrograph™ (DM).

Samples for TEM analysis were prepared by dispersing a small amount of $CrI_3$-MWCNT powder in absolute ethanol through mild sonication. The resulting dispersion was drop-cast onto lacey carbon-coated copper grids and left to dry under ambient conditions.

ASSOCIATED CONTENT

**Supporting Information**

Low-magnification HAADF-STEM images of encapsulated and coated $CrI_3$ nanotubes (Figure S3); in situ beam irradiation analysis of $CrI_3$ nanotube–nanorod junction (Figure S5); complete experimental details, supplementary figures, and additional discussions (PDF).

Video S1

Time-resolved in situ HRTEM video showing the nanotube-to-nanorod transformation of an encapsulated $CrI_3$ nanotube under electron beam exposure (MP4).

Video S2

In situ HRTEM video showing accelerated transformation dynamics of a $CrI_3$ nanotube–nanorod junction under electron beam irradiation (MP4).

AUTHOR INFORMATION

**Corresponding Author**

Ihsan Çaha and Francis Leonard Deepak - *International Iberian Nanotechnology Laboratory, 4715-330 Braga, Portugal*

E-mail: ihsan.caha@inl.int, leonard.francis@inl.int

**Author Contributions**

The manuscript was written through the contributions of all authors. All authors have given approval to the final version of the manuscript.

**Notes**

The authors declare no competing financial interest.

ACKNOWLEDGMENT

We acknowledge financial support from the European Union (Grant FUNLAYERS - 101079184)

REFERENCES

(1)    Ahn, E. C. 2D Materials for Spintronic Devices. *npj 2D Mater. Appl.* **2020**, *4* (1).

https://doi.org/10.1038/s41699-020-0152-0.

(2) Galbiati, M.; Zatko, V.; Godel, F.; Hirschauer, P.; Vecchiola, A.; Bouzehouane, K.; Collin, S.; Servet, B.; Cantarero, A.; Petroff, F.; Martin, M. B.; Dlubak, B.; Seneor, P. Very Long Term Stabilization of a 2D Magnet down to the Monolayer for Device Integration. *ACS Appl. Electron. Mater.* **2020**, *2* (11), 3508–3514. https://doi.org/10.1021/acsaelm.0c00810.

(3) Gati, E.; Inagaki, Y.; Kong, T.; Cava, R. J.; Furukawa, Y.; Canfield, P. C.; Bud'ko, S. L. Multiple Ferromagnetic Transitions and Structural Distortion in the van Der Waals Ferromagnet VI3 at Ambient and Finite Pressures. *Phys. Rev. B* **2019**, *100* (9), 94408. https://doi.org/10.1103/PhysRevB.100.094408.

(4) Huang, B.; Clark, G.; Navarro-Moratalla, E.; Klein, D. R.; Cheng, R.; Seyler, K. L.; Zhong, Di.; Schmidgall, E.; McGuire, M. A.; Cobden, D. H.; Yao, W.; Xiao, D.; Jarillo-Herrero, P.; Xu, X. Layer-Dependent Ferromagnetism in a van Der Waals Crystal down to the Monolayer Limit. *Nature* **2017**, *546* (7657), 270–273. https://doi.org/10.1038/nature22391.

(5) Liu, Z.; Guo, Y.; Chen, Z.; Gong, T.; Li, Y.; Niu, Y.; Cheng, Y.; Lu, H.; Deng, L.; Peng, B. Observation of Intrinsic Crystal Phase in Bare Few-Layer CrI3. *Nanophotonics* **2022**, *11* (19), 4409–4417. https://doi.org/10.1515/nanoph-2022-0246.

(6) Shcherbakov, D.; Stepanov, P.; Weber, D.; Wang, Y.; Hu, J.; Zhu, Y.; Watanabe, K.; Taniguchi, T.; Mao, Z.; Windl, W.; Goldberger, J.; Bockrath, M.; Lau, C. N. Raman Spectroscopy, Photocatalytic Degradation, and Stabilization of Atomically Thin Chromium Tri-Iodide. *Nano Lett.* **2018**, *18* (7), 4214–4219. https://doi.org/10.1021/acs.nanolett.8b01131.

(7) Gish, J. T.; Lebedev, D.; Stanev, T. K.; Jiang, S.; Georgopoulos, L.; Song, T. W.; Lim, G.; Garvey, E. S.; Valdman, L.; Balogun, O.; Sofer, Z.; Sangwan, V. K.; Stern, N. P.; Hersam, M. C. Ambient-Stable Two-Dimensional CrI3 via Organic-Inorganic Encapsulation. *ACS Nano* **2021**, *15* (6), 10659–10667. https://doi.org/10.1021/acsnano.1c03498.

(8) Lee, Y.; Choi, Y. W.; Lee, K.; Song, C.; Ercius, P.; Cohen, M. L.; Kim, K.; Zettl, A. 1D Magnetic MX3 Single-Chains (M = Cr, V and X = Cl, Br, I). *Adv. Mater.* **2023**, *35* (49), 1–8. https://doi.org/10.1002/adma.202307942.

(9) Li, Y.; Hu, Z.; Guo, Q.; Li, J.; Liu, S.; Xie, X.; Zhang, X.; Kang, L.; Li, Q. Van Der Waals One-Dimensional Atomic Crystal Heterostructure Derived from Carbon Nanotubes. *Chem. Soc. Rev.* **2025**. https://doi.org/10.1039/d4cs00670d.

(10) Cambré, S.; Liu, M.; Levshov, D.; Otsuka, K.; Maruyama, S.; Xiang, R. Nanotube-Based 1D Heterostructures Coupled by van Der Waals Forces. *Small* **2021**, *17* (38), 1–26. https://doi.org/10.1002/smll.202102585.

(11) Lu, S.; Guo, D.; Cheng, Z.; Guo, Y.; Wang, C.; Deng, J.; Bai, Y.; Tian, C.; Zhou, L.; Shi, Y.; He, J.; Ji, W.; Zhang, C. Controllable Dimensionality Conversion between 1D and 2D CrCl3 Magnetic Nanostructures. *Nat. Commun.* **2023**, *14* (1). https://doi.org/10.1038/s41467-023-38175-4.

(12) Cabana, L.; Ballesteros, B.; Batista, E.; Magén, C.; Arenal, R.; Orõ-Solé, J.; Rurali, R.; Tobias, G. Synthesis of PbI2 Single-Layered Inorganic Nanotubes Encapsulated within Carbon Nanotubes. *Adv. Mater.* **2014**, *26* (13), 2016–2021. https://doi.org/10.1002/adma.201305169.

(13) Anumol, E. A.; Deepak, F. L.; Enyashin, A. N. Capillary Filling of Carbon Nanotubes by BiCl3: TEM and MD Insight. *Nanosyst. Physics, Chem. Math.* **2018**, No. September, 521–531. https://doi.org/10.17586/2220-8054-2018-9-4-521-531.

(14) Batra, N. M.; Ashokkumar, A. E.; Smajic, J.; Enyashin, A. N.; Deepak, F. L.; Costa, P. M. F. J. Morphological Phase Diagram of Gadolinium Iodide Encapsulated in Carbon Nanotubes. *J. Phys. Chem. C* **2018**, *122* (43), 24967–24976. https://doi.org/10.1021/acs.jpcc.8b05342.

(15) Ashokkumar, A. E.; Enyashin, A. N.; Deepak, F. L. Single Walled BiI3 Nanotubes Encapsulated within Carbon Nanotubes. *Sci. Rep.* **2018**, *8* (1), 2–9. https://doi.org/10.1038/s41598-018-28446-2.

(16) Nakanishi, Y.; Furusawa, S.; Sato, Y.; Tanaka, T.; Yomogida, Y.; Yanagi, K.; Zhang, W.; Nakajo, H.; Aoki, S.; Kato, T.; Suenaga, K.; Miyata, Y. Structural Diversity of Single-Walled Transition Metal Dichalcogenide Nanotubes Grown via Template Reaction. *Adv. Mater.* **2023**, *35* (46). https://doi.org/10.1002/adma.202306631.

(17) Çaha, I.; Ahmad, A.; Boddapatti, L.; Bañobre-lópez, M.; Costa, A. T.; Enyashin, A. N.; Li, W.; Gargiani, P.; Valvidares, M.; Fernández-rossier, J.; Deepak, F. L. One-Dimensional CrI3 Encapsulated within Multi-Walled Carbon Nanotubes. *Commun. Chem.* **2025**, *8*, 155. https://doi.org/10.1038/s42004-025-01550-x.

(18) Zhang, T.; Grzeszczyk, M.; Li, J.; Yu, W.; Xu, H.; He, P.; Yang, L.; Qiu, Z.; Lin, H.; Yang, H.; Zeng, J.; Sun, T.; Li, Z.; Wu, J.; Lin, M.; Loh, K. P.; Su, C.; Novoselov, K. S.; Carvalho, A.; Koperski, M.; Lu, J. Degradation Chemistry and Kinetic Stabilization of Magnetic CrI3.

*J. Am. Chem. Soc.* **2022**, *144* (12), 5295–5303. https://doi.org/10.1021/jacs.1c08906.

(19) Sandoval, S.; Kepić, D.; Pérez Del Pino, Á.; György, E.; Gómez, A.; Pfannmoeller, M.; Tendeloo, G. Van; Ballesteros, B.; Tobias, G. Selective Laser-Assisted Synthesis of Tubular van Der Waals Heterostructures of Single-Layered PbI2 within Carbon Nanotubes Exhibiting Carrier Photogeneration. *ACS Nano* **2018**, *12* (7), 6648–6656. https://doi.org/10.1021/acsnano.8b01638.

(20) Flahaut, E.; Sloan, J.; Friedrichs, S.; Kirkland, A. I.; Coleman, K. S.; Williams, V. C.; Hanson, N.; Hutchison, J. L.; Green, M. L. H. Crystallization of 2H and 4H PbI2 in Carbon Nanotubes of Varying Diameters and Morphologies. *Chem. Mater.* **2006**, *18* (8), 2059–2069. https://doi.org/10.1021/cm0526056.

(21) Kreizman, R.; Hong, S. Y.; Sloan, J.; Popovitz-Biro, R.; Albu-Yaron, A.; Tobias, G.; Ballesteros, B.; Davis, B. G.; Green, M. L. H.; Tenne, R. Core-Shell PbI2@WS2 Inorganic Nanotubes from Capillary Wetting. *Angew. Chemie - Int. Ed.* **2009**, *48* (7), 1230–1233. https://doi.org/10.1002/anie.200803447.

(22) Botos, A.; Biskupek, J.; Chamberlain, T. W.; Rance, G. A.; Stoppiello, C. T.; Sloan, J.; Liu, Z.; Suenaga, K.; Kaiser, U.; Khlobystov, A. N. Carbon Nanotubes as Electrically Active Nanoreactors for Multi-Step Inorganic Synthesis: Sequential Transformations of Molecules to Nanoclusters and Nanoclusters to Nanoribbons. *J. Am. Chem. Soc.* **2016**, *138* (26), 8175–8183. https://doi.org/10.1021/jacs.6b03633.

(23) Sandoval, S.; Pach, E.; Ballesteros, B.; Tobias, G. Encapsulation of Two-Dimensional Materials inside Carbon Nanotubes: Towards an Enhanced Synthesis of Single-Layered

Metal Halides. *Carbon N. Y.* **2017**, *123*, 129–134. https://doi.org/10.1016/j.carbon.2017.07.031.

(24) Yomogida, Y.; Nagano, M.; Liu, Z.; Ueji, K.; Rahman, M. A.; Ahad, A.; Ihara, A.; Nishidome, H.; Yagi, T.; Nakanishi, Y.; Miyata, Y.; Yanagi, K. Semiconducting Transition Metal Dichalcogenide Heteronanotubes with Controlled Outer-Wall Structures. *Nano Lett.* **2023**, *23* (22), 10103–10109. https://doi.org/10.1021/acs.nanolett.3c01761.

(25) Hu, C.; Michaud-Rioux, V.; Yao, W.; Guo, H. Theoretical Design of Topological Heteronanotubes. *Nano Lett.* **2019**, *19* (6), 4146–4150. https://doi.org/10.1021/acs.nanolett.9b01661.

(26) Xiang, R.; Inoue, T.; Zheng, Y.; Kumamoto, A.; Qian, Y.; Sato, Y.; Liu, M.; Tang, D.; Gokhale, D.; Guo, J.; Hisama, K.; Yotsumoto, S.; Ogamoto, T.; Arai, H.; Kobayashi, Y.; Zhang, H.; Hou, B.; Anisimov, A.; Maruyama, M.; Miyata, Y.; Okada, S.; Chiashi, S.; Li, Y.; Kong, J.; Kauppinen, E. I.; Ikuhara, Y.; Suenaga, K.; Maruyama, S. One-Dimensional van Der Waals Heterostructures. *Science (80-. ).* **2020**, *367* (January), 537–542.

(27) Kuklin, A. V.; Visotin, M. A.; Baek, W.; Avramov, P. V. CrI3 Magnetic Nanotubes: A Comparative DFT and DFT+U Study, and Strain Effect. *Phys. E Low-Dimensional Syst. Nanostructures* **2020**, *123* (May), 114205. https://doi.org/10.1016/j.physe.2020.114205.

(28) S. Djurdjić-Mijin, A. Šolajić, J. Pešić, M. Šćepanović, Y. Liu, A. Baum, C. Petrovic, N. Lazarević, Z. V. P. Lattice Dynamics and Phase Transition in CrI3 Single Crystals. *Phys. Rev. B* **2018**, *98* (10), 104307. https://doi.org/10.1103/PhysRevB.98.104307.

(29) Telkhozhayeva, M.; Girshevitz, O. Roadmap toward Controlled Ion Beam-Induced Defects

in 2D Materials. *Adv. Funct. Mater.* **2024**, *2404615*, 1–31. https://doi.org/10.1002/adfm.202404615.

(30) Costa, P. M. F. J.; Gautam, U. K.; Bando, Y.; Golberg, D. Direct Imaging of Joule Heating Dynamics and Temperature Profiling inside a Carbon Nanotube Interconnect. *Nat. Commun.* **2011**, *2* (1). https://doi.org/10.1038/ncomms1429.

(31) Zhang, R.; Feng, Y.; Li, H.; Kumamoto, A.; Wang, S.; Zheng, Y.; Dai, W.; Fang, N.; Liu, M.; Tanaka, T.; Kato, Y. K.; Kataura, H.; Ikuhara, Y.; Maruyama, S.; Xiang, R. Fabricating One-Dimensional van Der Waals Heterostructures on Chirality-Sorted Single-Walled Carbon Nanotubes. *Carbon N. Y.* **2022**, *199* (August), 407–414. https://doi.org/10.1016/j.carbon.2022.07.076.

(32) You, A.; Be, M.; In, I. Nanocomposite under Electron Irradiation. *AIP Conf. Procedings* **2008**, *999* (April), 79–92.